\documentclass[aps,prx,twocolumn,superscriptaddress]{revtex4-2}
\usepackage{amsmath}
\usepackage{amssymb}
\usepackage{color}
\usepackage{graphicx}
\usepackage{epstopdf}
\usepackage{natbib}

\usepackage{multirow}
\begin{document}
	
	\title{Enhanced zeta potentials caused by surface ion mobilities }
	\author{Evgeny S. Asmolov}
	\affiliation{Frumkin Institute of Physical Chemistry and Electrochemistry, Russian Academy of Sciences, 31 Leninsky Prospect, 119071 Moscow, Russia}
	\affiliation{Institute of Mechanics, Lomonosov Moscow State
		University, 1 Michurinskiy Prospect, 119991 Moscow, Russia}
	\author{Elena F. Silkina}
	\affiliation{Frumkin Institute of Physical Chemistry and Electrochemistry, Russian Academy of Sciences, 31 Leninsky Prospect, 119071 Moscow, Russia}
	\author{Olga I. Vinogradova}
	\email[Corresponding author: ]{oivinograd@yahoo.com}
	\affiliation{Frumkin Institute of Physical Chemistry and Electrochemistry, Russian Academy of Sciences, 31 Leninsky Prospect, 119071 Moscow, Russia}
	\begin{abstract}
		The electro-hydrodynamics near conducting walls is revisited. Attention is focused on the impact of an explicit diffuse Stern layer, which
		permittivity and viscosity  differ from the bulk values, on the velocity of an electro-osmotic plug flow. To solve this problem we propose an approach of mapping the flow in the Stern layer to  the surface dividing the Stern and diffuse layer, where an effective electro-hydrodynamic slip boundary condition is imposed. The latter implies that  an effective surface charge is  responding to the applied field and characterized by a mobility parameter $\mu \geq 1$. We derive analytic equations for $\mu$ and demonstrate that it is determined only by electrostatic properties of the electric double layer.
		These equations are then used to calculate electrokinetic (zeta) potentials of surfaces. We show that the zeta potential generally exceeds the surface one, which implies an amplification of the electro-osmotic flow. This effect is most pronounced if the hydrodynamic slip length is large and/or in concentrated solutions.
		
	\end{abstract}
	
	\maketitle
	
	%\title{Electrokinetic mobility of the Stern layer caused by permittivity and viscosity contrasts}
	
	\affiliation{Frumkin Institute of Physical Chemistry and Electrochemistry
		Russian Academy of Sciences, 31-4 Leninsky Prospect, 119071 Moscow, Russia}
	
	\affiliation{Frumkin Institute of Physical Chemistry and Electrochemistry
		Russian Academy of Sciences, 31-4 Leninsky Prospect, 119071 Moscow, Russia}
	
	\affiliation{Frumkin Institute of Physical Chemistry and Electrochemistry
		Russian Academy of Sciences, 31-4 Leninsky Prospect, 119071 Moscow, Russia}
	
	\affiliation{Frumkin Institute of Physical Chemistry and Electrochemistry
		Russian Academy of Sciences, 31-4 Leninsky Prospect, 119071 Moscow, Russia}
	
	\affiliation{Frumkin Institute of Physical Chemistry and Electrochemistry
		Russian Academy of Sciences, 31-4 Leninsky Prospect, 119071 Moscow, Russia}
	
	%\affiliation{Frumkin Institute of Physical Chemistry and   Electrochemistry, Russian Academy of Science, 31 Leninsky Prospect,   119071 Moscow, Russia}
	
	\section{Introduction}
	
	When an  electric field $E$ is applied tangent to a charged wall, an electro-osmotic flow of an electrolyte solution is induced~\cite{anderson.jl:1989}. This phenomenon, termed electroosmosis, takes its origin in the adjacent electric double layer. The classical model assumes that the latter includes an inner Stern layer of a thickness $\delta$ below a few molecular sizes and a so-called  outer electrostatic diffuse layer (EDL) that extends to distances of the order of the Debye
	length $\lambda$ of a bulk electrolyte solution~\cite{stern.o:1924}.  Since long time it was axiomatic in colloid science that the voltage drop in this inner layer is caused by a specific (i.e. non-Coulombic) adsorption of potential-determining ions~\cite{lyklema:2010}, and such chemisorbed ions are traditionally treated as immobile. This layer can be considered as forming part of the wall, so the (effective) surface potential $\Phi_s$ is  traditionally defined at the plane separating inner and outer regions of the double layer. Note that in the old textbooks on colloid science it is often termed a ``slip plane''~\cite{note1}. In 1921 Smoluchowski presented an elegant theory of an electroosmotic flow by postulating the immobile (stagnant) Stern layer,
	and argued that the finite velocity $U_{\infty}$ outside the EDL is given by~\cite{smoluchowski.m:1921}
	\begin{equation}\label{eq:smoluchowsky}
		U_{\infty} = - \dfrac{\varepsilon E}{4 \pi \eta } Z,
	\end{equation}
	with permittivity of the solution $\varepsilon$, its dynamic
	viscosity $\eta$, and so-called (electro-hydrodynamic) zeta-potential  $Z$ of the surface, where the no-slip boundary condition, $U_s = 0$, is postulated. This postulate implies, via the Stokes equation, that $Z$ must be equal to $\Phi_s$, or, equivalently, to the drop of the electrostatic potential within the diffuse layer.
	
	Since Smoluchowski, colloid scientists have traditionally considered that the EDL only is electrokinetically active by excluding the Stern layer from consideration.
	This is, in fact, unnecessary, if the no-slip boundary condition is applied. Indeed, since  immobile chemisorbed ions cannot set liquid to motion, the inner flow could emerge only if the liquid velocity on the surface does not vanish. However, the picture can become more sophisticated, if the tangential liquid velocity is different
	from that of the solid surface, \emph{i.e.} $U_s \neq 0$. For instance, at the hydrophobic surface~\cite{vinogradova.oi:1999}
	\begin{equation}\label{eq:slip0}
		U_{s}=b \partial _{z} U,
	\end{equation}
	where $b$ is the slip length and $\partial _{z} U$ is the surface shear rate. In this case a liquid flow inside the Stern layer may appear, \emph{i.e.} it will no longer be stagnant. The inner flow can be excluded from the analysis of the outer electro-osmotic flow if surface ions are fixed and there are no inner mobile (diffuse) ions. However, Eq.~\eqref{eq:slip0} cannot be justified for a liquid-gas interface (bubbles or
	drops, foams) as well as for some hydrophobic solids. To describe the electro-osmotic fluid velocity in these cases, an electro-hydrodynamic boundary condition has been formulated~\cite{maduar.sr:2015}%
	\begin{equation}
		U_{s}=b\left[ \partial _{z} U+\frac{(1-\mu )\sigma E}{\eta }\right],
		\label{bc_dim}
	\end{equation}%
	where $\sigma $ is a surface charge density, and $\mu $ is a parameter accounting for a mobility of surface charges that has  been  ignored until
	recently. Such a mobility, however, exists, is supported by simulation data~\cite{maduar.sr:2015,grosjean.b:2019}, but has immense variability
	depending
	on substrate material~\cite{mangaud.e:2022}.
	
	\begin{figure}[h]
		%\hspace{-3.25cm}
		\centering
		\includegraphics[width=0.9\columnwidth ]{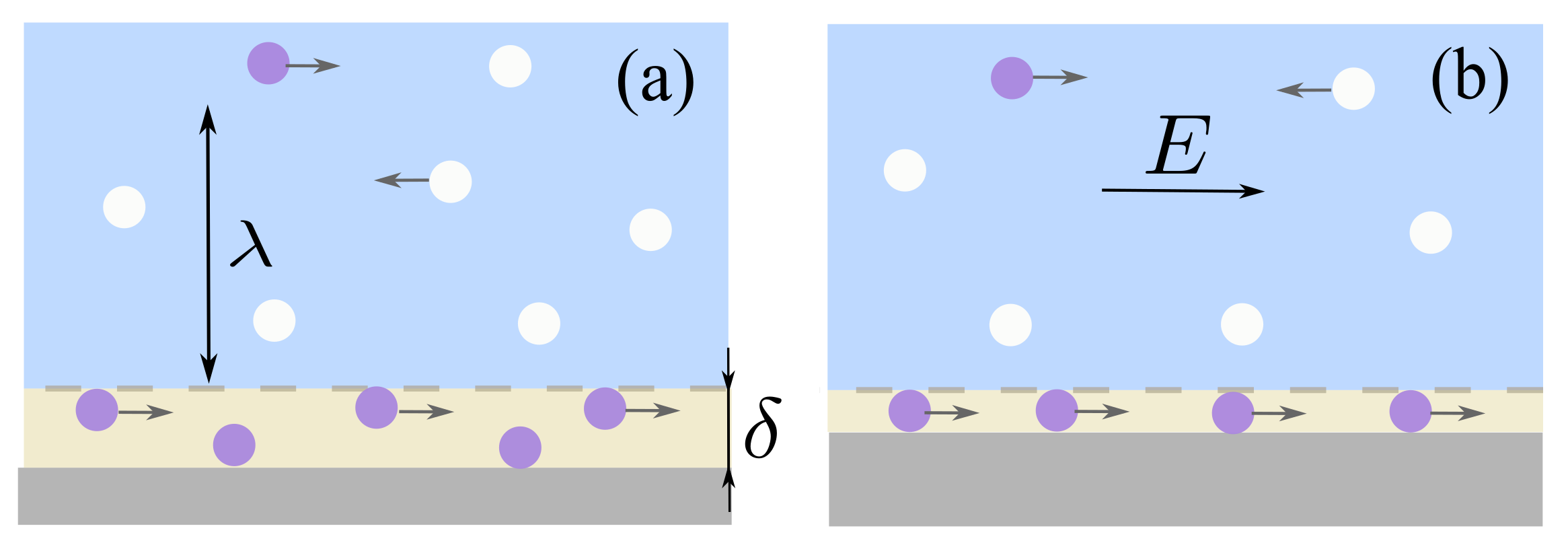}
		\caption{Schematic representation of two mechanisms of  mobility of surface ions leading to a backward inner flow. Cations and anions are denoted with colored and white circles.}
		\label{fig:mu_0}
	\end{figure}
	
	During the last years several theoretical papers have been concerned with the interpretation of $\mu$~\cite{vinogradova.oi:2023}. It now seems certain that there is not one, but at least two surface ion mobility mechanisms. One~\cite{maduar.sr:2015} relies on the ``gas cushion model'', where the inner region is described as a lubricating film of a reduced viscosity~\cite{vinogradova.oi:1995a}. In this model $b$ depends on $\delta$ and viscosity contrast, and $\mu$ is treated as the fraction of ions, which are specifically chemisorbed at a solid wall, as shown in Fig.~\ref{fig:mu_0}(a). The surface ions (of fraction $1-\mu $) are considered as mobile and frictionless, which implies that in a pressure-driven flow, they
	translate with the velocity of a hydrodynamic slip $U_{s}$ given by Eq.~\eqref{eq:slip0}, but do not
	migrate relative to liquid.
	Another model~\cite{mouterde.t:2018} assumes that the slip length is given and that the inner layer is composed by physisorbed potential-determining (frictional) ions. This relates $\mu$ to a momentum portion that they transfer to the surface under an applied electric or pressure field [see Fig.~\ref{fig:mu_0}(b)]. Both mechanisms imply that the ions of the Stern layer, which are excluded from thermal motion, and counter-ions of the EDL are oppositely charged. Thus by
	migrating in response to the electric field the inner ions drag fluid in the direction opposite to the main (outer) electroosmotic flow. As a result the attainable $\mu$ is bounded and can vary from 0 to 1. When $\mu = 0$  the electroosmosis is suppressed and the zeta potential vanishes, while if $\mu = 1$, Eq.~\eqref{bc_dim} reduces to \eqref{eq:slip0}. Thus, both these mechanisms imply that the mobility of surface ions, if any, can only reduce the flow and could be  seen as detrimental effect.
	Some recent developments, however, suggest another line to attack the problem of the mobility of surface ions.
	
	The first of these has to do with the computer simulations~\cite{maduar.sr:2015,grosjean.b:2019} that have revealed that
	the Stern layer contains spatially distributed diffuse ions being in thermodynamic equilibrium with the bulk solution. Although the thickness of
	this layer is only about one to two ion diameters, the (lateral) mobility remains large~\cite{mouterde.t:2019}.
	
	The second development has to do with the permittivity of the solvent near the wall. The concept of drastically reduced permittivity in the Stern layer has been introduced to interpret the double layer capacitance~\cite
	{conway.be:1951}, is consistent with the later theories~\cite{borisevich.sv:1999,guidelli.r:2000} and computer simulations~\cite{
		hartkamp.r:2018,schlaich.a:2016,becker.m:2023}, and is supported by modern capacitance experiments~\cite{fumagalli.l:2018}. Besides, the surface
	potentials of conductors inferred from direct surface force measurements
	are well consistent with a theory based on such a model of the Stern layer~\cite{vinogradova.oi:2024a}.

	If the Stern layer is of reduced permittivity and contains the diffuse ions, the implication for an electro-osmotic flow may be large. The point is that the inner ions would be of the same sign as in the outer EDL, which implies that they now would drag fluid in the direction of the outer flow and augment $\mu$ and, consequently, $Z$. That possibility to enhance the electro-osmotic flow has been completely ignored so far. Our purpose in this paper is to derive analytic expressions for $\mu$ and $Z$ of conductors using the physically based simple model~\cite{bonthuis.dj:2012,uematsu.y:2018}, which assumes that  the permittivity  and fluid viscosity in the Stern layer remain constant  (and reduced compared to the bulk).
	
	Our paper is arranged as follows: In Sec.~\ref{sec:model} we define our model system and present governing equations. Section~\ref{sec:theory} describes the concept of effective surface charge and the procedure for calculating $\mu$. We derive the general expression for $\mu$ and argue that it reflects solely the
	electrostatic properties of the electric double layer. We then present approximate formulas for different modes.
	In Sec.~\ref{sec:theory_zeta} we focus attention on the zeta potential and flow amplification that emerges due to a nonzero slip velocity at the surface. We conclude in Sec.~\ref{sec:conclusion}.
	
	\section{Model}\label{sec:model}
	
\begin{figure}[h]
		%\hspace{-3.25cm}
		\centering
		\includegraphics[width=1\columnwidth ]{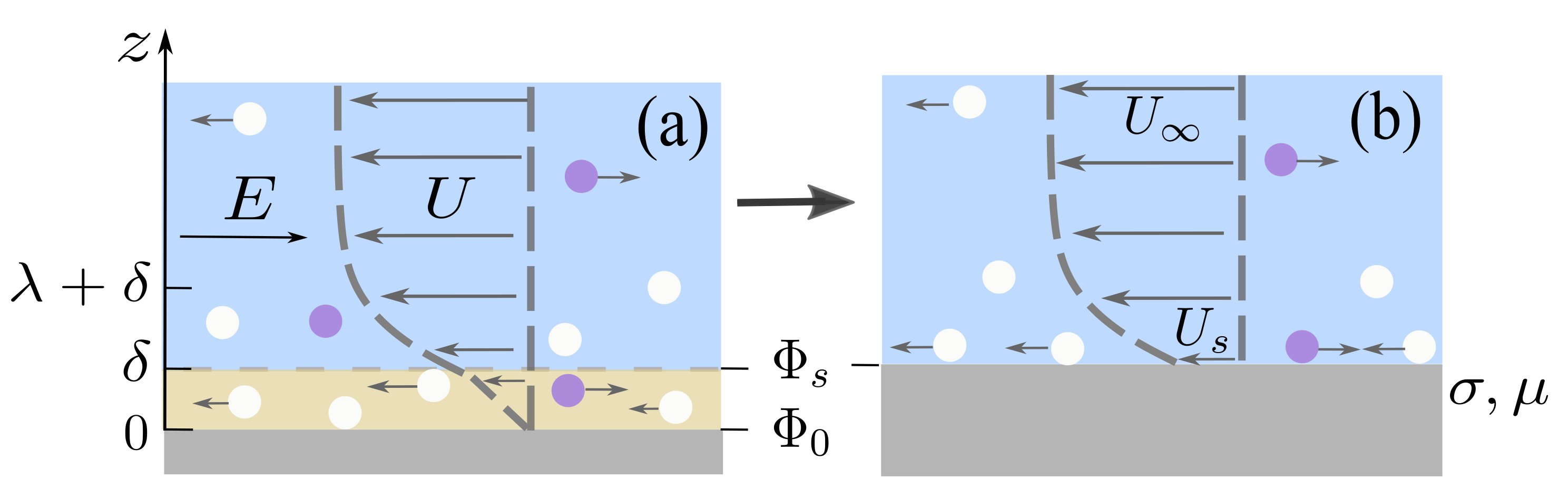}
		\caption{The real double layer (a) and the imaginary wall (b) of the effective charge density $\sigma$ and mobility parameter $\mu$. }
		\label{fig:1}
	\end{figure}

	We consider a planar wall of a fixed potential $\Phi_0$ located at $z=0$ and unbounded in the $x$ and $y$ directions [see Fig.~\ref{fig:1}(a)]. The wall is in contact with a reservoir of  1:1 salt solution at temperature $T$ and number density $n_{\infty}$. The Debye length of a bulk solution, $\lambda=\left( 8\pi \ell _{B}n_{\infty}\right) ^{-1/2}$, is defined as usually with the Bjerrum
	length, $\ell _{B}=\dfrac{e^{2}}{\varepsilon k_{B}T}$, where $e$ is the proton charge, $k_{B}$ is the Boltzmann
	constant, and $\varepsilon$ is the solvent permittivity.  By analyzing the experimental data it is more convenient to use the concentration $c_{\infty}[\rm{mol/l}]$, which is related to $n_{\infty} [\rm{m^{-3}}] $
	as $n_{\infty} \simeq N_A \times 10^3 \times c_{\infty}$, where $N_A$ is Avogadro's number. The Bjerrum
	length of water at $T \simeq 298$ K is equal to about $0.7$ nm leading to a
	useful formula for 1:1 electrolyte
	\begin{equation}\label{eq:DLength}
		\lambda [\rm{nm}] \simeq \frac{0.305 [\rm{nm}]}{\sqrt{c_{\infty}[\rm{mol/l}]} }
	\end{equation}
	Thus upon increasing $c_{\infty}$ from $10^{-6}$ to $10^{-1}$ mol/l the screening length is reduced from about 300 down to 1 nm.
	
	The electrolyte ions occupy the positive half
	plane. Near a wall the double layer, which includes an adjacent to a body inner Stern layer and an outer electrostatic diffuse layer, is formed. The inner region is defined at $0\leq z\leq \delta $, an outer region at $z\geq \delta $, and $\lambda
	_{o}\equiv \lambda $. It is supposed that in the outer layer the permeability and dynamic viscosity are isotropic and have the same values as in the bulk of the liquid, i.e. $\varepsilon_o = \varepsilon$ and $\eta_o = \eta$, where index \emph{o} denotes an outer region. Inside the Stern layer the permittivity $\varepsilon_i \ll \varepsilon$ and $\eta_i \neq \eta$, where \emph{i} denotes an inner region.
	
	There are still some uncertainties about what value to assign to $\delta$ and $\gamma =\varepsilon /\varepsilon_i$. Since inorganic ions have hydrodynamic diameters from 0.2 to 0.6  nm~\cite{kadhim.mj:2020} we might argue that  $\delta$  cannot be smaller. The best fit on experimental data  in pure water yielded $\delta \simeq 0.75 \pm 0.15$ nm~\cite{fumagalli.l:2018}. In our calculations below we will employ $\delta = 0.5$ nm, which has been successfully used before to interpret the measurements of the capacity of the inner layer~\cite{conway.be:1999,kornyshev.aa:1982} and of the surface potential~\cite{vinogradova.oi:2024a}.
	The (normal) permittivity in the Stern layer has been earlier assumed to have a value of 4~\cite{conway.be:1951}. Consequent indirect experiments  suggested that $\varepsilon_i$ varies rather from 6 to 30~\cite{butt.hj:2003}, which corresponds to a reduction of  $\gamma$ from \emph{ca.} $13$ down to 3. However, we here employ $\gamma \simeq 40$ that has been measured in a recent experiment~\cite{fumagalli.l:2018}. Note that this value of $\gamma \gg 1$ has been successfully used to interpret the surface force measurements~\cite{vinogradova.oi:2024a}.
	
	Ions in both layers obey Boltzmann distribution, $c_{\pm }(z)=c_{\infty}\exp (\mp \phi_{i,o} (z))$, where $\phi_{i,o} (z)=e\Phi_{i,o}(z)/(k_{B}T)$
	[recall that $k_{B}T/e \simeq 25$ mV] is the dimensionless electrostatic
	potential,  the upper (lower) sign
	corresponds to cations (anions).

	The electrostatic potential satisfies the nonlinear Poisson-Boltzmann equation~\cite{herrero.c:2024}:%
	\begin{equation}
		\phi _{i,o}^{\prime \prime }=\lambda _{i,o}^{-2}\sinh \phi _{i,o}.
		\label{eq:PB2}
	\end{equation}%
	where $^{\prime }$ denotes $d/dz$. The inner Debye length $\lambda _{i}=\left( 8\pi \ell
	_{i}n_{\infty }\right) ^{-1/2}$, where $\ell _{i}=\dfrac{e^{2}}{\varepsilon
		_{i}k_{B}T}$. In what follows $\ell_i = \gamma \ell_B$ and $\lambda ^{2}=\gamma \lambda _{i}^{2}$. Clearly, the inner Debye
	length is smaller than the outer one or equal to it.
	
	The boundary condition at the conducting
	wall is that of a constant potential
	\begin{equation}
		\phi_i (0)=\phi _{0}.
	\end{equation}%
	We remark that $\phi _{0}$ can be quite large, so here we will use $\phi
	_{0}\leq 20$ that corresponds to 0.5 V.
	
	The potential at the surface is continuous,%
	\begin{equation}
		\phi _{s}=\phi _{i}(\delta )=\phi _{o}(\delta ),
	\end{equation}%
	but the gradient of potential changes across the surface as
	\begin{equation}
		\phi _{i}^{\prime }(\delta )=\gamma \phi _{o}^{\prime }(\delta ).
	\end{equation}
	
	%The boundary condition at the channel wall can also be expressed in terms
	%of the surface charge,
	%\begin{equation}
	%\phi _{i}^{\prime }(0)=-\dfrac{4\pi \sigma _{0}\ell _{i}}{e}.
	%\end{equation}%

	%Whereas  the charge density of the real wall is $\sigma_0$,  the effective $\sigma$ for an imaginary wall (surface) has an extra contribution that represents an integrated (per unit area) charge of diffuse ions   accumulated inside the
	%Stern layer:
	%\begin{equation}
	%\sigma _{i}=-en_{\infty }\int_{0}^{\delta }\sinh \phi _{i}dz.  \label{si}
	%\end{equation}
	%In other words, the standard term is supplemented by the term due to finite diffuse Stern layer: $\sigma = \sigma_0 + \sigma_i$.

	%Since fluid flow can be induced by hydrodynamic and electrostatic forces the slip velocity at the surface includes two contributions. One of them is proportional
	%to the shear rate, another - to the electric field.

	The flow satisfies the Stokes equation with an electrostatic body force
	\begin{equation}
		\partial _{z}\left[ \eta_{i,o} \partial _{z} U_{i,o} \right] =-\rho_{i,o} E,
		\label{eq:Stokes}
	\end{equation}%
	where $U_o = U$ and $\rho_{i,o} =-2en_{\infty }\sinh \phi_{i,o} $ is the
	volume charge density. We calculate the
	contribution of interfacial layer to the slip velocity for flow induced by
	electric field. The boundary condition at the wall reads%
	\begin{equation}
		U_i(0)=0.  \label{bc0}
	\end{equation}%
	At the surface
	\begin{equation}\label{eq:BC_surface}
		U_{s}=U_{i}(\delta )=U(\delta ),
	\end{equation}
	and far away from the wall ($z \to \infty$),%
	\begin{equation}
		U = U_{\infty}; \, U' = 0.  \label{bch}
	\end{equation}

	\section{Effective surface charge and its mobility}\label{sec:theory}
	
	Macroscopically, the defined in Sec.~\ref{sec:model} double-layer system would appear as an imaginary wall located at $z=\delta$ of an effective surface charge density $\sigma$, which  satisfies
	\begin{equation}
		\phi _{o}^{\prime }(\delta )=-\dfrac{4\pi \sigma\ell _{B}}{e}.
		\label{eq:BC_charge}
	\end{equation}
	This is illustrated in Fig.~\ref{fig:1}(b).
	Boundary condition~\eqref{bc_dim} can then be used as an effective one that is applied at the surface and mimics the actual two-layer system in the outer region. This effective condition fully characterizes the outer flow generated by the real wall and can be used to solve the electroosmotic problem without tedious calculations.
	
	%\subsection{Mobility parameter}\label{sec:theory_mu}

	First integration of the Stokes equation~\eqref{eq:Stokes} from an arbitrary $z$ to $\infty$ and applying boundary condition (\ref{bch}) yields the following shear rates in the inner and outer layers
	\begin{equation}
		U'_{i,o} = \frac{E}{\eta_{i,o} }\int_{z}^{\infty }\rho_{i,o}\left( s\right) ds.
		\label{dvzeo}
	\end{equation}
	
	A further definite integration imposing no-slip boundary condition (\ref{bc0}) at the wall  gives the velocity profiles:
	\begin{equation}
		U_{i,o}=E\int_{0}^{z}\int_{t}^{\infty }\frac{\rho_{i,o} \left( s\right) }{\eta_{i,o} \left(
			t\right) }dsdt.  \label{vzeo}
	\end{equation}
	
	From Eq.~\eqref{eq:BC_surface} it follows then that the slip velocity at the surface is given by
	\begin{equation}\label{us_eo_tot}
		U_{s}=E\int_{0}^{\delta }\int_{t}^{\infty }\frac{\rho_{i,o} \left( z\right) }{\eta
			_{i}}dzdt.
	\end{equation}
	It is convenient to re-express Eq.~\eqref{us_eo_tot} in the form  identical to electro-hydrodynamic boundary condition~(\ref{bc_dim}) by dividing $U_s$ into an outer and inner layer contribution
	\begin{equation}
		U_s = \frac{E}{\eta _{i}}\left[ \delta \int_{\delta }^{\infty }\rho_{o} \left( z\right)
		dz+\int_{0}^{\delta }\int_{t}^{\delta }\rho_{i} \left( z\right) dzdt\right] .
		\label{us_eo}
	\end{equation}

	The first term in \eqref{us_eo} is proportional to the shear rate at the surface [given by Eq.~\eqref{dvzeo}] and associated with the contribution of the EDL. This coincides  with the first term in (\ref{bc_dim}) provided the slip
	length is given by~\cite{vinogradova.oi:1995a}%
	\begin{equation}
		b = \delta \frac{\eta }{\eta _{i}}.  \label{bcm}
	\end{equation}
	This equation is identical to derived within the ``gas cushion model''~\cite{vinogradova.oi:1995a,vinogradova.oi:1995c}, but recall that in the present case the surface is defined at $z=\delta$ and $\eta_i \neq \infty$. This dictates that the slip length is now positive even if $\eta_i \geq \eta$. For instance, $b = \delta$, if the viscosity is assumed to be the same in the Stern layer and in the bulk solution.
	The upper attainable value of $b/\delta\simeq 50$ corresponds to a gas layer.
	
	One further comment should be made. It follows from Eq.~\eqref{bcm} that $b$ is a hydrodynamic length scale that is decoupled with  electrostatics. This conclusion is derived for the situation displayed in Fig.~\ref{fig:mu_0}(a). Another, illustrated in Fig.~\ref{fig:mu_0}(b), model is different: it, in principle, suggests that the slip length is affected by the surface charge. Note, however, that computer simulations relying on this model suggest that electrostatic contribution to friction, if any, is extremely small and can be neglected~\cite{joly.l:2006b,sendner.c:2009,xie.y:2020}.
	
	The second terms in (\ref{us_eo}) is associated with the contribution of the Stern layer. By changing the integration order one can easily transform the double integral
	\begin{equation} \label{emuI}
		\mathcal{P} = \int_{0}^{\delta }\int_{t}^{\delta }\rho_i
		\left( z\right) dzdt.
	\end{equation}
	into single%
	\begin{equation}
		\mathcal{P} =\int_{0}^{\delta }\rho_i \left( z\right)
		\int_{0}^{z}dtdz=\int_{0}^{\delta }z\rho_i \left( z\right) dz.  \label{int1}
	\end{equation}%
	Since $U_i(z) \propto z$, the quantity $z\rho_i \left( z\right)$ can be identified with the momentum of ions located at a distance $z$ from the wall (per unit volume). Thus,  $\mathcal{P}$   characterizes a total momentum (per unit area) of diffuse ions confined inside the inner layer of thickness $\delta$.
	
	The second terms in (\ref{bc_dim}) and \eqref{us_eo} coincide if
	\begin{equation}
		\frac{b (1-\mu )\sigma}{\eta }=\frac{ \mathcal{P}}{\eta _{i}}.  \label{emu}
	\end{equation}%
	It  follows then that
	\begin{equation}
		\sigma (1 - \mu) = \frac{ \mathcal{P}}{\delta}.  \label{mu1}
	\end{equation}
	Thus the value of $\sigma (1 - \mu)$ is related to the averaged (over volume) momentum of ions in the Stern layer that is induced by an applied
	field.
	Our treatment clarifies the status of $\mu$. The limiting case of special interest is that for which $\mu = 0$.
	It follows from \eqref{mu1} that this particular situation implies that $\mathcal{P}=\sigma \delta $. This may occur when all mobile ions are located at $z=\delta^-$. Substituting $\rho _{i}=\sigma \delta _{D}\left( z-\delta^-\right)$, where $\delta _{D}$ is the Dirac delta-function, in Eq.~\eqref{int1} and taking the integral we obtain that $\mathcal{P}$ is indeed a product of $\sigma$ and $\delta$. Another case of special interest is that of $\mu =1$, which corresponds to a stagnant Stern layer. If so,
	the left-hand side in Eq.~\eqref{mu1} vanishes and, consequently, $\mathcal{P}=0$. Such a scenario may occur when inner ions are all located at the wall, $\rho _{i}=\sigma \delta _{D}\left(z^+\right)$, as can easily be proven by substituting this into \eqref{int1} and performing integration.
	
	We remark that to interpret non-integer values of $\mu$
	it is not necessary to introduce fractions of mobile and immobile ions into the problem  [as in Fig.~\ref{fig:mu_0}(a)], nor it is necessary to make specific assumptions about  friction of inner ions on liquid and wall [as in Fig.~\ref{fig:mu_0}(b)]. The latter appear  naturally and are
	established self-consistently, if the inner ions are diffuse. Indeed,  from Eq.~\eqref{mu1} it follows that the values of $\mu$ confined from 0 to 1 are attained when $\mathcal{P}$ and $\sigma$ are of the same sign. In our case, however, $\mu$  should be above unity or equal to it since $\mathcal{P}$ is either of the opposite sign to that of $\sigma$ or vanishes.

	We are now in a position to make a connection between $\mu$ and the double layer properties.
	Integrating Eq.~(\ref{int1}) and using (\ref{eq:PB2}) we obtain
	\begin{eqnarray}
		\mathcal{P}
		&=&-2en_{\infty }\lambda _{i}^{2}\int_{0}^{\delta }\left[ \phi _{i}^{\prime
		}\left( \delta \right) -\phi_{i} ^{\prime }\left( z\right) \right] dz
		\label{ibox} \\
		&=&-2en_{\infty }\lambda _{i}^{2}\left[ \delta \phi _{i}^{\prime }\left(
		\delta \right) - \Delta\phi\right],  \label{ib2}
	\end{eqnarray}%
	where $\Delta\phi = \phi_0 - \phi_s$ is the potential drop in the Stern layer.
	The first term in Eq.~(\ref{ib2}) can then be re-expressed in terms of $\sigma$
	using (\ref{eq:BC_charge}):
	\begin{equation*}
		-2en_{\infty }\lambda _{i}^{2}\delta \phi _{i}^{\prime }\left( \delta
		\right) =-2en_{\infty }\lambda ^{2}\delta \phi _{o}^{\prime }\left( \delta
		\right) =\sigma\delta ,
	\end{equation*}%
	so that (\ref{mu1}) becomes
	\begin{equation}\label{eq:mu_inter}
		\mu =\frac{2en_{\infty }\lambda _{i}^{2}\Delta\phi }{%
			\delta \sigma}=\frac{e\Delta\phi }{4\pi \ell
			_{i}\delta \sigma}.
	\end{equation}%
	To eliminate $\sigma$ from this equation one can invoke the Grahame relation between the surface charge and potential:
	\begin{equation}
		\sigma=\frac{e}{2\pi \lambda \ell _{B}}\sinh \left( \frac{\phi _{s}}{2}\right).
		\label{grahame}
	\end{equation}%
	Substituting Eq.~\eqref{grahame} into Eq.~\eqref{eq:mu_inter}
	we obtain
	\begin{equation}
		\mu =\frac{\Delta\phi}{2\sinh \left( \frac{\phi _{s}}{2}\right) }\frac{\lambda }{\gamma \delta }.  \label{eq:muf}
	\end{equation}%
	In the case of small and large surface potentials this general expression for $\mu$ can be recast to simpler formulas.
	For small surface potential, $\phi_s \leq 1$, Eq.~\eqref{eq:muf} can be expanded about $\phi_s = 0$ and to first order it reduces to
	\begin{equation}\label{eq:small_phis}
		\mu \simeq \left( \frac{\phi _{0}}{\phi _{s}}-1\right) \frac{\lambda }{\gamma \delta }.
	\end{equation}
	However, for $\phi_s \geq 4$ one can safely use $2\sinh \left( \frac{\phi _{s}}{2}\right) \simeq \exp \left( \frac{\phi _{s}}{2}\right) $, so that Eq.~\eqref{eq:muf} can be simplified to
	\begin{equation}\label{eq:large_phis}
		\mu \simeq \frac{\Delta\phi}{\exp \left( \frac{\phi _{s}}{2}\right) }\frac{%
			\lambda }{\gamma \delta }
	\end{equation}
	The magnitude of $\mu$ does not depend on the viscosity contrast, but depends on the established self-consistently $\phi _{s}$ and the potential drop in the Stern layer. Other parameters that control $\mu$ are $\gamma \gg 1$ and $\lambda/\delta$. Thus, we can conclude that $\mu$  reflects the static properties of the electric double layer solely. Once $\phi_s$ is determined, Eq.~\eqref{eq:muf} or its approximations \eqref{eq:small_phis} and \eqref{eq:large_phis} provide a direct route to  calculating $\mu$.
	
	\begin{figure}[h]
		%\hspace{-4.15cm}
		\centering
		\includegraphics[width=0.6\columnwidth ]{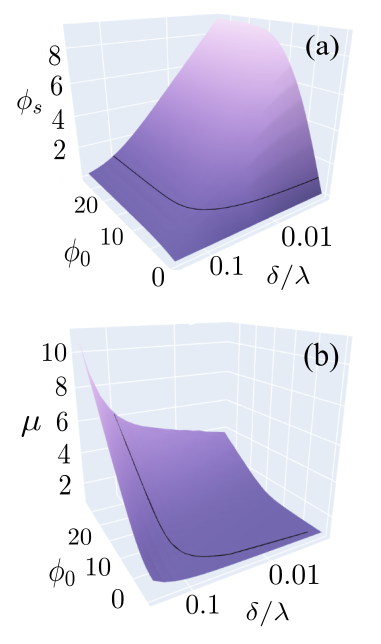}
		%Colorbar  shows the values of $\phi_s$, and the region of $\phi_s \leq 1$ is colored by gray.}
	\caption{Surface potential $\phi_s$ (a) and parameter $\mu$ (b) as a function of $\phi_0$ and $\delta/\lambda$ computed using $\gamma = 40$. A contour line connects the points, where $\phi_s = 1$. }
	\label{fig:3d}
\end{figure}

The surface potential, in turn, may be found numerically from~\cite{vinogradova.oi:2024a}
\begin{equation}  \label{eq:final_int}
	\int_{\phi_s}^{\phi_0}\dfrac{d\phi_i}{\sqrt{2\gamma \left[\cosh \phi_{i} +
			(\gamma - 1) \cosh \phi_s -\gamma\right]}} = \frac{ \delta}{ \lambda},
\end{equation}
which is exact and applicable for any $\phi_0$, as well as being very well suited for numerical work. In Fig.~\ref{fig:3d}(a) we plot $\phi_s$ computed from Eq.~\eqref{eq:final_int} for $\gamma = 40$  as a function of two variables, $\phi_0$ and $\delta/\lambda$.
It can be seen that $\phi_s$ is extremely sensitive  to $\delta/\lambda$ and generally depends on $\phi_0$. The surface potential augments on increasing $\phi_0$ when the latter is small enough, but for sufficiently large $\phi_0$ it attains its upper possible value and becomes insensitive to further increase in the intrinsic potential. A contour line that corresponds to $\phi_s = 1$ is included to indicate a bound of a region of small surface potentials. We see that $\phi_s$ becomes small only if $\phi_0\leq1$ or at high salt.

The results for $\mu$ computed using $\gamma = 40$ are shown in Fig.~\ref{fig:3d}(b). An overall conclusion from this three dimensional plot is that $\mu$ increases with $\phi_0$ and $\delta/\lambda$.

Some useful approximate solutions for $\phi_s$ are known and can be immediately used to derive approximate expressions for $\mu$ in different modes. Below we
consider two distinct situations,  of small and large $\phi_0$.

We focus first to the case of small applied potentials,  $\phi_0 \leq 1$. The (small) $\phi_s$ is
given by~\cite{vinogradova.oi:2024a}
\begin{equation}\label{eq:phis_small_general}
	\phi_s \simeq  \dfrac{\phi_0}{\cosh \left(\frac{\sqrt{ \gamma}}{ \lambda} \delta \right) + \sqrt{ \gamma} \sinh \left(\frac{\sqrt{ \gamma}}{ \lambda} \delta \right)}.
\end{equation}
%Thus in this limit $\phi_s$ increases linearly with $\phi_0$. The slope of this line is defined by $\delta/\lambda$ and also depends on $\sqrt{\gamma}$
From Eq.~\eqref{eq:small_phis} it follows that
\begin{equation}\label{eq:mu_DH}
	\mu \simeq \left(\cosh \left(\frac{\sqrt{ \gamma}}{ \lambda} \delta \right) + \sqrt{ \gamma} \sinh \left(\frac{\sqrt{ \gamma}}{ \lambda} \delta \right)-1\right) \frac{\lambda }{\gamma \delta }
\end{equation}
does not depend on $\phi_0$ and $\phi_s$. This can further be divided into two limits.

In the limit of $\sqrt{\gamma}\delta/\lambda \ll 1$, which is the case of dilute solutions, the surface potential is given by
\begin{equation}\label{eq:uematsu}
	\phi_s \simeq \dfrac{\phi_0}{1 + \dfrac{\gamma \delta}{ \lambda} },
\end{equation}
and
Eq.~\eqref{eq:mu_DH} reduces to
\begin{equation}\label{eq:mu=1}
	\mu \simeq 1
\end{equation}
indicating that the Stern layer remains immobile.

For $\sqrt{\gamma}\delta/\lambda \ge 1$ that refers to concentrated solutions,
\begin{equation}\label{eq:phis_small_general2}
	\phi_s \simeq \dfrac{2 \phi_0 e^{-\sqrt{ \gamma} \delta/\lambda}}{1 + \sqrt{ \gamma}},
\end{equation}
and we derive
\begin{equation}\label{eq:mu0}
	\mu \simeq  \left(\frac{(1 + \sqrt{\gamma})e^{\sqrt{\gamma}\delta/\lambda}}{2} - 1 \right) \frac{\lambda}{\gamma \delta}.
\end{equation}
Thus, $\mu$ decays with $\lambda \leq \sqrt{\gamma} \delta$ (increases with $c_{\infty}$).

As another example we consider the situation, when
$\phi_0$ is large. In this case two limiting scenarios occur depending on the value of $\phi_s$.

For dilute solutions $\phi_s$ becomes large. It first augments nonlinearly on increasing $\phi_0$ and then saturates to~\cite{vinogradova.oi:2024a}
\begin{equation}\label{eq:plateau}
	\phi_{s} \simeq \ln \left( \dfrac{\lambda^2 \ln^2 [4(\gamma - 1)] }{\delta^2 \gamma (\gamma - 1)} \right).
\end{equation}
Note that from Eq.~\eqref{eq:DLength} it follows that $\lambda^2 \propto c_{\infty}^{-1}$, which suggests that $\phi_s \propto -\ln(c_{\infty})$.
Substituting \eqref{eq:plateau} into \eqref{eq:large_phis} then yields
\begin{equation}\label{eq:mu_largephi}
	\mu \simeq \frac{\Delta\phi \sqrt{\gamma (\gamma - 1)}}{\gamma\ln [4(\gamma - 1)] }.
\end{equation}
Thus, $\mu \propto \Delta\phi$, but in reality, $\mu$ grows linearly with $\phi_0$ (since $\phi_s$ remains constant). If $\phi_0$ is fixed, but we vary (low) $c_{\infty}$, then an increment in $\mu$ grows weakly logarithmically with $c_{\infty}$.

If $\phi_s \leq 1$, which is typical for concentrated solutions, it can be approximated by~\cite{vinogradova.oi:2024a}:
\begin{equation}\label{eq:small_pot}
	\phi _{s}\simeq\frac{8}{1+\sqrt{\gamma }} e^{-\frac{\sqrt{%
				\gamma }}{\lambda }\delta },
\end{equation}
and approximate expression \eqref{eq:small_phis} for $\mu$ can be written as
\begin{equation}\label{eq:large0smalls}
	\mu \simeq  \left(\frac{\phi_0(1 + \sqrt{\gamma})e^{\sqrt{\gamma}\delta/\lambda}}{8} - 1 \right) \frac{\lambda}{\gamma \delta},
\end{equation}
so in this mode $\mu \propto \phi_0$.

It is of interest to compare numerical data with the approximate theory and to determine the regimes of validity of these approximate results. Below we present results based on  numerical solutions of Eqs.~\eqref{eq:PB2} and~\eqref{eq:Stokes} with prescribed boundary conditions together with specific calculations using above asymptotic approximations.

\begin{figure}[h]
	%\hspace{-3.25cm}
	\centering
	\includegraphics[width=0.99\columnwidth ]{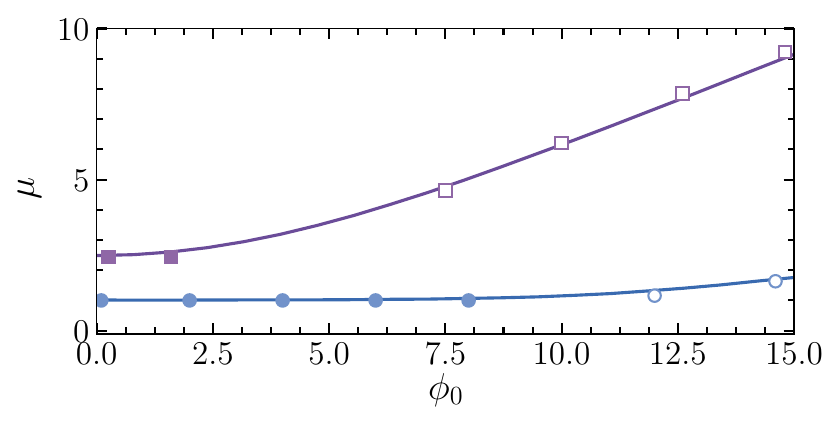}
	\caption{The parameter $\mu$ as a function of $\phi_0$
		computed using $\delta = 0.5$ nm and $\gamma = 40$ for $c_{\infty} = 5 \times 10^{-1}$ (upper curve) and $10^{-5}$ mol/l (lower curve). Filled and open circles show calculations from Eqs.~\eqref{eq:mu=1} and ~\eqref{eq:mu_largephi}. Filled and open squares correspond to Eqs.~\eqref{eq:mu0} and \eqref{eq:large0smalls}.  }
	\label{fig:mu_small}
\end{figure}

We begin by studying the dependence of $\mu$ on $\phi_0$. The values of $\delta = 0.5$ nm and $\gamma = 40$ are fixed, and we compute $\mu$ for two different $c_{\infty} = 10^{-5}$ and $5\times10^{-2}$ mol/l that correspond to $\lambda$ ca. 100 and 1.4 nm. The results are shown in Fig.~\ref{fig:mu_small}. It can be seen that for smaller $\phi_0$ on both curves the branches of constant $\mu$ occur, but at larger applied potentials $\mu$ increases strictly  monotonically.
For an upper curve ($c_{\infty} = 5\times10^{-2}$ mol/l and $\sqrt{\gamma}\delta/\lambda \simeq 2.3 \geq 1$) the plateau branch extends up to $\phi_0 \simeq 1$ and is well fitted by Eq.~\eqref{eq:mu0}. It is interesting to note that $\mu$ is above unity when $\phi_0$ vanishes.
For a lower curve ($c_{\infty} = 10^{-5}$ mol/l and $\sqrt{\gamma}\delta/\lambda \simeq 0.03 \ll 1$) this branch extends to $\phi_0 \simeq 8$ and is described by Eq.~\eqref{eq:mu=1} derived for small applied potentials. This equation thus has validity beyond the range its formal applicability and $\mu$ can be equal to unity also in the intermediate mode, where $\phi_s$ is rather large and grows non-linearly with $\phi_0$.
Also included are calculations from Eqs.~\eqref{eq:mu_largephi} and \eqref{eq:large0smalls} derived for a large $\phi_0$ mode, where $\phi_s$ saturates. One can conclude that these approximate expressions fit numerical results very well.

\begin{figure}[h]
	%\hspace{-4.15cm}
	\centering
	\includegraphics[width=0.99\columnwidth ]{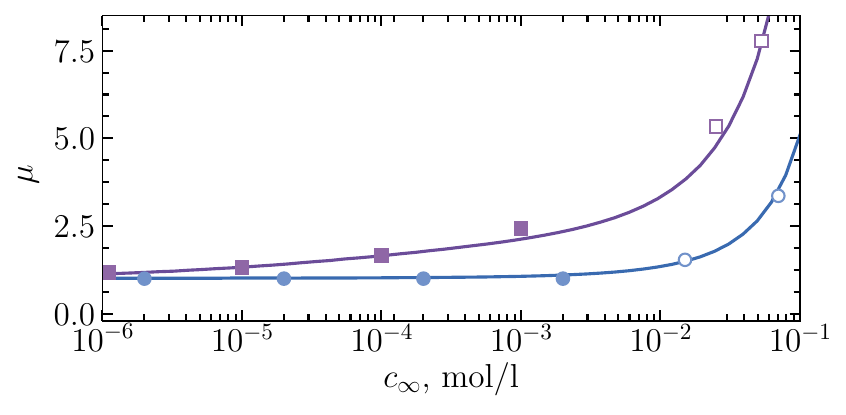}
	\caption{The parameter $\mu$ \emph{vs.} $c_{\infty}$ computed
		using $\gamma = 40$ and $\delta = 0.5$ nm for $\phi_0 = 12$ (upper curve) and $\phi_0 = 2$ (lower curve). Filled and open circles correspond to Eqs.~\eqref{eq:mu=1} and \eqref{eq:mu0}. Filled and open squares are calculated using Eqs.~\eqref{eq:mu_largephi} and \eqref{eq:large0smalls}.  }
	\label{fig:salt}
\end{figure}

The concentration dependence of $\mu$ is of interest. Let us now fix $\phi_0$ and vary  $c_{\infty}$ using the same $\delta$ and $\gamma$ as before. For these numerical calculations, made using $\phi_0 = 2$ and 12, we increase $c_{\infty}$ from $10^{-6}$ to $10^{-2}$ mol/l.
The results shown in Fig.~\ref{fig:salt} demonstrate that $\mu$ indeed responds to salt concentration. For $\phi_0 = 2$ we observe that Eq.~\eqref{eq:mu=1} applies up to $c_{\infty} \simeq 2\times10^{-2}$  mol/l.
For very dilute electrolyte solutions and $\phi_0 =2$, we observe $\mu \simeq 1$.
For a larger value of $c_{\infty}$, the mobility
increases monotonically. The transition between modes, when $\mu$ becomes greater than unity, occurs at concentrations $2\times10^{-3}$ and $2\times10^{
	-2}$ mol/l for surfaces with $\phi_0 = 12$ and 2.

\section{Zeta potential and flow amplification}\label{sec:theory_zeta}

We turn now to the (dimensionless) zeta potential $\zeta = e Z/(k_B T)$ of a single wall, which can be calculated as~\cite{silkina.ef:2019}:
\begin{equation}\label{eq:eo}
	\zeta = \phi_{s} - u_s = \phi_{s} + \frac{2\mu b}{\lambda}\sinh \left( \frac{\phi _{s}}{2}\right),
\end{equation}
where $u_s = \displaystyle\frac{4\pi \ell _{B}\eta }{e E} U_s$ is the dimensionless slip velocity at the surface.
The presence of diffuse ions in the Stern layer should have significant repercussions for $\zeta$ due to emergence of $u_s$. Since $\mu \geq 1$ the zeta potential may potentially augment even for hydrophilic surfaces, where $b/\delta \leq 1$.

Note that from Eq.~\eqref{eq:muf} it follows that
Eq.~\eqref{eq:eo} is equivalent to
\begin{equation}\label{eq:eo_final}
	\zeta \simeq  \phi_s + \frac{\Delta\phi}{\gamma}\frac{ b}{ \delta },
\end{equation}
which provides an alternative route to the determination of zeta potentials [Eq.~\eqref{eq:eo_final} was also derived by~\citet{uematsu.y:2018} by using different arguments, which confirms the validity of our approach].

In concentrated solutions $\lambda$ becomes very small, so is $\phi_s$. Consequently, $\zeta \simeq -u_s$. In other words at high concentrations nearly all diffuse ions are confined inside the Stern layer and generate the slip velocity at the surface, which is approximately equal to the outer velocity (\emph{i.e.} in the EDL and the bulk). This immediately leads to
\begin{equation}\label{eq:zeta_large2_LB}
	\zeta \simeq \dfrac{b}{\delta}\dfrac{ \phi_0} {\gamma}.
\end{equation}
This equation specifies the lowest bound on attainable $\zeta$. The latter can become quite high if both $b$ and $\phi_0$ are large. Note that nearly vanishing $\phi_s$ is equivalent to  $\sigma \simeq 0$. Since $\mathcal{P}\neq 0$, from Eq.~\eqref{mu1} it follows that  $\mu \to \infty$.
In highly dilute solutions $\lambda$ is large and $\phi_s$ can also becomes large, although bounded.

The amplification factor  that characterizes an enhancement of an outer plug flow relative to
what is generated near surfaces of $\zeta = \phi_s$
can be evaluated as
\begin{equation}\label{eq:A_definition}
	\mathcal{A} = \frac{\zeta}{\phi_s} = 1 - \frac{u_s}{\phi_s}.
\end{equation}
Using \eqref{eq:eo} and \eqref{eq:eo_final} this gives
\begin{equation}\label{eq:A}
	\mathcal{A} =  1 + \frac{2\mu b}{\lambda \phi_s}\sinh \left( \frac{\phi _{s}}{2}\right) \equiv 1 + \frac{\Delta\phi}{\phi_s} \frac{b}{\gamma \delta }
\end{equation}
Either first or second equality can be used for specific calculations depending on the mathematical convenience.

%\subsection*{Small surface potential}

The above general equations for $\zeta$ and $\mathcal{A}$ include $\phi_s$, which, in turn, is controlled by several system parameters. In order to obtain detailed information concerning the zeta potential and amplification factor we have to use the specific expressions for the surface potential.
Below we  describe the results of calculations  in the limits of small and large $\phi_0$.

We begin by studying the situation of small $\phi_0$. In this case $\phi_s$ and $\mu$ are given by Eqs.~\eqref{eq:phis_small_general} and \eqref{eq:mu_DH}, and it is convenient to use Eq.~\eqref{eq:eo}, which for small surface potentials  reduces to
\begin{equation}\label{eq:eo_small}
	\zeta \simeq  \phi_{s}  + \frac{\mu b}{\lambda}\phi_s.
\end{equation}

For dilute solutions, where the argument $\frac{\sqrt{\gamma }}{\lambda }\delta $ becomes small leading to Eq.~\eqref{eq:mu=1} for $\mu$, and  \eqref{eq:eo_small} yields a well-known expression~\cite{joly.l:2004}
\begin{equation}\label{eq:eo_small1}
	\zeta \simeq \phi_s \left(1 + \frac{b}{\lambda}\right),
\end{equation}
which leads to
\begin{equation}\label{eq:eo_small1_A}
	\mathcal{A} \simeq 1 + \frac{b}{\lambda}.
\end{equation}
Taking into account the order
of magnitudes, one can speculate that  in dilute solutions the flow amplification should be small, if not negligible. Later we shall see that this is indeed so.

For more concentrated solutions, $\frac{\sqrt{\gamma }}{\lambda }\delta \geq 1$, the standard calculations give $\mu \phi_s \simeq  \frac{\phi_0 \lambda}{\gamma \delta}$, which yields
\begin{equation}\label{eq:eo_small2}
	\zeta \simeq \phi_s + \frac{b \phi_0}{\gamma \delta} \simeq \phi_0 \left(\dfrac{2  e^{-\sqrt{ \gamma} \delta/\lambda}}{1 + \sqrt{ \gamma}} + \frac{b}{\gamma \delta}\right)
\end{equation}
Obviously, when $\lambda$ becomes very small, this equation reduces to \eqref{eq:zeta_large2_LB}.
The amplification factor can be then approximated by
\begin{equation}\label{eq:A_small2}
	\mathcal{A} \simeq  1 + \dfrac{(1 + \sqrt{\gamma}) b e^{\sqrt{ \gamma} \delta/\lambda}}{2\delta\gamma}.
\end{equation}
On reducing $\lambda$ (increasing $c_{\infty}$), the second term in \eqref{eq:A_small2} grows exponentially, so that in concentrated solutions $\mathcal{A}$ may become quite large, especially when $b/\delta \gg 1$.

A similar analysis can be carried our for a situation of a large intrinsic potential. We mention below only the case of $\phi_0 \geq 4$, where $\phi_s$ and $\mu$ are approximated by Eqs.~\eqref{eq:small_pot} and \eqref{eq:large0smalls}. Substituting these into \eqref{eq:eo_small} we find
\begin{equation}\label{eq:zeta_large2}
	\zeta \simeq  \left(1-\dfrac{b}{\delta \gamma}\right)  \frac{8}{1+\sqrt{\gamma }} e^{-\frac{\sqrt{\gamma }}{\lambda }\delta }
	+ \dfrac{b}{\delta \gamma} \phi_0.
\end{equation}
Simple estimates show that when $\frac{\sqrt{\gamma }\delta}{\lambda } \ge 1$, the second term dominates and Eq.~\eqref{eq:zeta_large2_LB} is reproduced.
Then dividing Eq.~\eqref{eq:zeta_large2} by \eqref{eq:small_pot} we obtain
\begin{equation}\label{eq:A_large2}
	\mathcal{A} \simeq  1 + \phi_0  \dfrac{(1+\sqrt{\gamma})be^{\frac{\sqrt{\gamma }}{\lambda }\delta }}{8\delta\gamma} .
\end{equation}

We now turn to a situation of sufficiently large $\phi_s$. If the plateau mode is reached, which can be attained at $\phi_0 \gg 1$ and low concentrations, the surface potential is given by \eqref{eq:plateau}. Using \eqref{eq:eo_final} we then find
\begin{equation}\label{eq:zeta_large1}
	\zeta \simeq  \left(1-\dfrac{b}{\delta \gamma}\right) \ln \left( \dfrac{\lambda^2 \ln^2 [4(\gamma - 1)] }{\delta^2 \gamma (\gamma - 1)} \right) + \dfrac{b}{\delta \gamma} \phi_0,
\end{equation}
and
\begin{equation}\label{eq:A_large1}
	\mathcal{A} \simeq  1 + \dfrac{b}{\delta \gamma} \left[\phi_0  \ln \left( \dfrac{\delta^2 \gamma (\gamma - 1)}{\lambda^2 \ln^2 [4(\gamma - 1)]} - 1\right)\right].
\end{equation}

\begin{figure}[h]
	%\hspace{-4.15cm}
	\centering
	\includegraphics[width=0.99\columnwidth ]{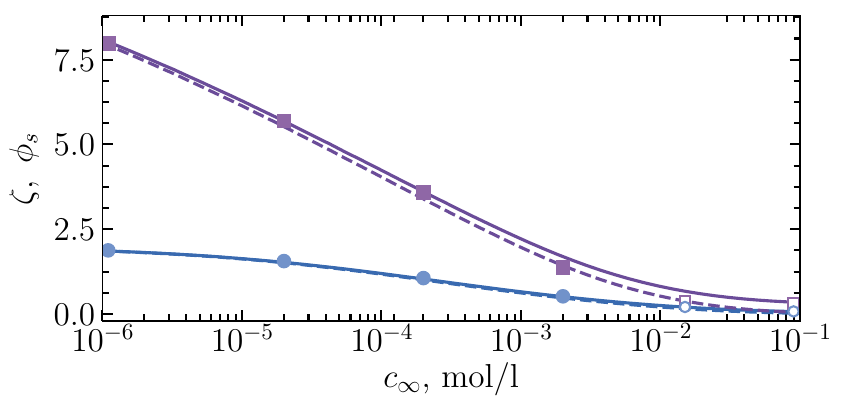}
	\caption{Zeta (solid) and surface (dashed) potentials as a function of $c_{\infty}$ computed
		using $\gamma = 40$ and $\delta = b = 0.5$ nm for $\protect\phi_0 = 12$ (upper set of curves) and $\phi_0 = 2$ (lower set of curves). Filled and
		open circles are calculations from Eqs.~\eqref{eq:eo_small1} and \eqref{eq:eo_small2}. Filled and open squares correspond to Eqs.~\eqref{eq:zeta_large1} and \eqref{eq:zeta_large2}.}
	\label{fig:zeta1}
\end{figure}

Figure~\ref{fig:zeta1} is intended to illustrate the salt dependence of zeta and surface potentials in the case of equal inner and outer viscosities, $\eta_i = \eta$. Such a situation typically occurs near a  hydrophilic wall~\cite{vinogradova.oi:1999}. An explicit Stern layer implies that the slip length is defined at the surface. Then it follows from \eqref{bcm} that for a hydrophilic wall we should set $b=\delta$. Calculations made with this value of $b$ and the same parameters as in Fig.~\ref{fig:salt} show that at lower concentrations $\zeta$ is larger and reduces on increasing $c_{\infty}$. The numerical data are compared with the analytical approximations. It is seen that when $c_{\infty}\leq 2 \times 10^{-3}$ mol/l the curves obtained for $\phi_0 = 2$ and $12$ are well fitted by Eqs.~~\eqref{eq:eo_small1} and \eqref{eq:eo_small2}, correspondingly.
Also includes are the curves for $\phi_s$. It can be seen that the zeta and surface potential are fairly close, but in the case of large $\phi_0$ there is some small discrepancy that is growing with the concentration of salt.

\begin{figure}[h]
	%\hspace{-4.15cm}
	\centering
	\includegraphics[width=0.99\columnwidth ]{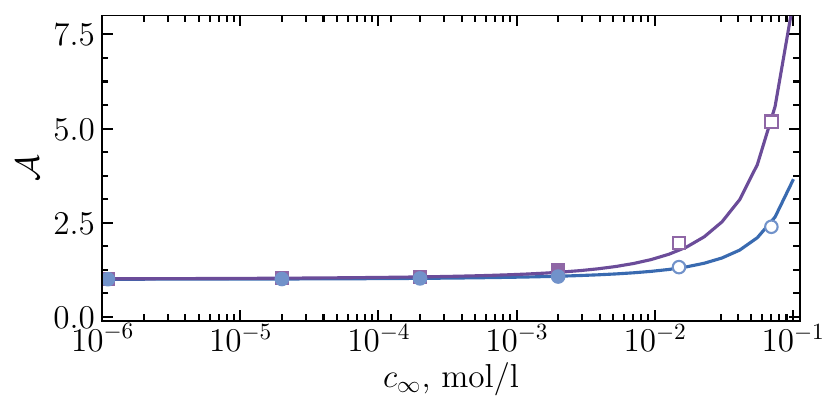}
	\caption{The amplification factor as a function of $c_{\infty}$ computed
		using $\gamma = 40$ and $\delta = b = 0.5$ nm for $\phi_0 = 12$ (upper curve) and $\phi_0 = 2$ (lower curve). Filled and open circles are calculations from Eqs.~\eqref{eq:eo_small1_A} and \eqref{eq:A_small2}. Filled and open squares correspond to Eqs.~\eqref{eq:A_large1} and \eqref{eq:A_large2}.}
	\label{fig:A1}
\end{figure}

To examine deviations of $\zeta$ from $\phi_s$ more closely, in Fig.~\ref{fig:A1} we plot (in lin-log scale) the amplification factor. Up to $c_{\infty} \simeq 2 \times 10^{-3}$ mol/l the curves for $\phi_0 = 2$ and 12 are well described by Eqs.~\eqref{eq:eo_small1_A} and \eqref{eq:A_small2}, but practically $\mathcal{A}\simeq 1$. If concentrations exceed $c_{\infty} \simeq 10^{-2}$ mol/l the exponential growth of $\mathcal{A}$ with salt is observed, and $\zeta$ becomes a few times larger than $\phi_s$. For $\phi_0 = 2$ this branch of the amplification curve is well fitted by Eq.~\eqref{eq:A_large2}, but if $\phi_0 \gg 1$,  Eq.~\eqref{eq:A_large1} applies. In any event, it seems clear that although such an amplification exists, it is unimportant. Since $\phi_s$ nearly vanishes, $\zeta$ remains small [see Fig.~\ref{fig:zeta1}]. An overall conclusion is thus a finite $\mu$ cannot generate enhanced electro-osmotic flow near hydrophilic surface.

\begin{figure}[h]
	%\hspace{-4.15cm}
	\centering
	\includegraphics[width=0.99\columnwidth ]{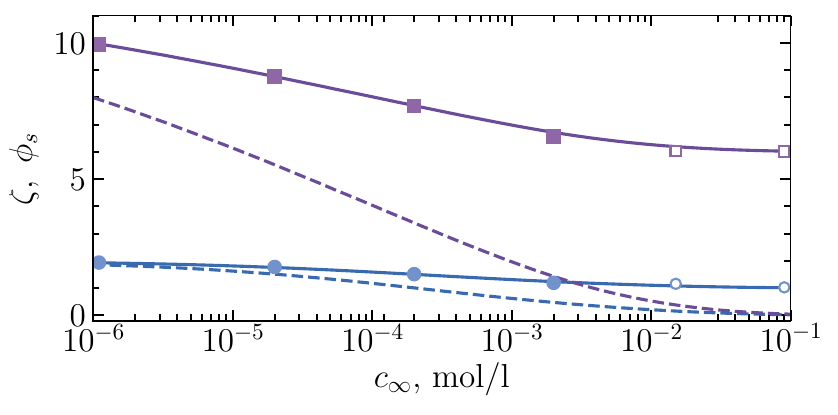}
	\caption{Zeta (solid) and surface (dashed) potentials as a function of $c_{\infty}$ computed
		using $\gamma = 40$ and $\delta = 0.5$ nm, $b = 10$ nm for $\phi_0 = 12$ (upper set of curves) and $\phi_0 = 2$ (lower set of curves). Filled and open circles are calculations from Eqs.~\eqref{eq:eo_small1} and \eqref{eq:eo_small2}. Filled and open squares correspond to Eqs.~\eqref{eq:zeta_large1} and \eqref{eq:zeta_large2}.}
	\label{fig:zeta}
\end{figure}

The hydrophobic surface is different since the inner layer viscosity is much smaller than $\eta$~\cite{vinogradova.oi:1995a}, which provides $b/\delta \gg 1$. The zeta potentials computed as a function of $c_{\infty}$ are shown in Fig.~\ref{fig:zeta}. The calculations are made using the same parameters as in Fig.~\ref{fig:zeta1}, so the surface potentials are reproduced from this figure, but now $b = 10$ nm is invoked. This has the effect of much larger zeta potentials, and the deviations from $\phi_s$ become significant, especially at high concentrations and large $\phi_0$. For $\phi_0 = 2$ the surface potential can be seen as rather small at all salt concentrations. Indeed, the numerical curve is well fitted by Eq.~\eqref{eq:eo_small1} at low concentrations and by \eqref{eq:eo_small2} at higher salt. For $\phi_0 = 12$ the low concentration branch corresponds to rather large  surface potentials, so we use Eq.~\eqref{eq:zeta_large1}  to fit the numerical data. Note that from Eq.~\eqref{eq:DLength} it follows that $\lambda^2 \propto c_{\infty}^{-1}$, which suggests that $\zeta \propto -\ln(c_{\infty})$. This is exactly what is observed in Fig.~\ref{fig:zeta}. When $c_{\infty}$ is above $10^{-2}$ mol/l the surface potential becomes small and we see that $\zeta$ is well described by \eqref{eq:zeta_large2}. Actually, at $c_{\infty} \simeq 10^{-1}$ mol/l the zeta potential already saturates to $ \zeta \simeq 6$ predicted by Eq.~\eqref{eq:zeta_large2_LB} and cannot become smaller.

\begin{figure}[h]
	%\hspace{-4.15cm}
	\centering
	\includegraphics[width=0.99\columnwidth ]{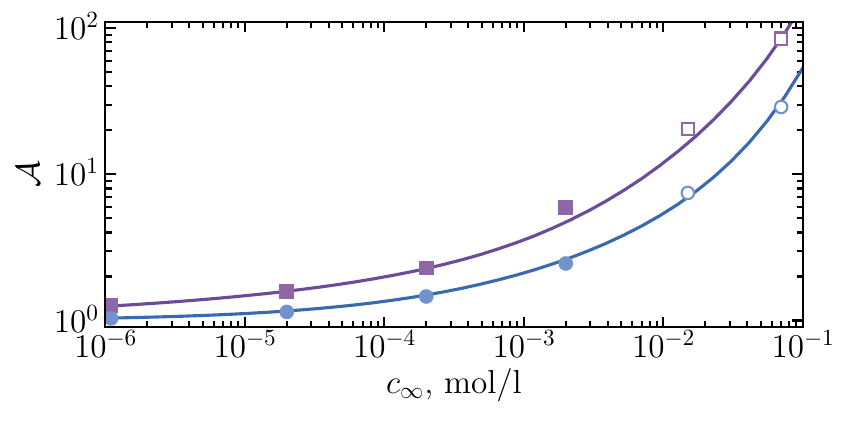}
	\caption{The amplification factor as a function of $c_{\infty}$ computed
		using $\gamma = 40$ and $\delta = 0.5$ nm, $b = 10$ nm for $\phi_0 = 12$ (upper curve) and $\phi_0 = 2$ (lower curve). Filled and open circles are calculations from Eqs.~\eqref{eq:eo_small1_A} and \eqref{eq:A_large1}. Filled and open squares correspond to Eqs.~\eqref{eq:A_small2} and \eqref{eq:A_large2}.}
	\label{fig:A}
\end{figure}

The salt dependence of $\mathcal{A}$ in this case of $b = 10$ nm is illustrated in Fig.~\ref{fig:A}. The results are now plotted in a log-log scale. We see that the implications of hydrodynamic slippage are large. Near hydrophobic walls the amplification factor is dramatically enhanced compared with the hydrophilic walls [cf. Fig.~\ref{fig:A1}], and $\mathcal{A}$ can be quite large even in dilute solutions. Finally, we mention that calculations made from Eqs.~\eqref{eq:eo_small1_A} and \eqref{eq:A_large1} perfectly fit the numerical data obtained using $\phi_0 = 2$. The data for $\phi_0 = 12$ are well fitted by  Eqs.~\eqref{eq:A_small2} and \eqref{eq:A_large2}.

\section{Conclusion}\label{sec:conclusion}

As emphasized in the introduction, this article has concentrated on the electro-osmotic
properties of a conductor wall that  is of a constant
potential. The new ingredient is the existence of explicit Stern layer of reduced permittivity that contain diffuse ions.
We have shown that such an inner layer participates in the flow-driving mechanism by reacting to the field. Since the inner ions are of the same sign as the outer electrostatic diffuse layer, they induce a flow in the same direction. Effectively, the Stern layer acts as a surface with adsorbed mobile charges, which obeys electro-hydrodynamic boundary condition~\eqref{bc_dim}, but note the difference from reported before mechanisms for this mobility. We recall that one involves fractions of chemisorbed and frictionless ions~\cite{maduar.sr:2015}. Another has to do with the momentum transfer of physisorbed frictional ions~\cite{mouterde.t:2018}. Both induce a backward flow, reducing the
amplification of electro-osmosis caused by a hydrodynamic slip~\cite{silkina.ef:2019}. By contrast, the forward inner flow generated within our mechanism can lead to a massive amplification of electro-osmosis at the slippery surface, especially in concentrated solutions.

The forward electro-osmotic flow inside the Stern layer is analogous to that inside charged porous coatings~\cite{silkina.ef:2020b,vinogradova.oi:2020}, but its origin and consequences are different. Inside the coating of a given (volume) charge density such a flow emerges since it absorbs diffuse counter-ions. As a result, large amplification factor can be attained in dilute solutions only, while here $\mathcal{A}$ augments with the concentration by showing a very rapid growth  at high salt.

\begin{table}[h]
\caption{Summary of equations to calculating $\phi_s$, $\mu$, and $\zeta$ in different modes.}\label{tab:summary1}
\begin{tabular}{|c|cc|cc|}
\hline
  $\phi_0$                     & \multicolumn{2}{c|}{$\phi_0\leq1$}    & \multicolumn{2}{c|}{$\phi_0\gg1$}    \\ \hline
\multirow{2}{*}{$c_{\infty}$}      & \multicolumn{1}{c|}{dilute} & concentrated & \multicolumn{1}{c|}{dilute} & concentrated \\
                       & \multicolumn{1}{c|}{$\sqrt{\gamma}\delta/\lambda \ll 1$} & $\sqrt{\gamma}\delta/\lambda \geq 1$ & \multicolumn{1}{c|}{$\phi_s \gg 1$} & $\phi_s \leq 1$ \\ \hline
            $\phi_{s}$           & \multicolumn{1}{c|}{Eq.~\eqref{eq:uematsu}} & Eq.~\eqref{eq:phis_small_general2} & \multicolumn{1}{c|}{Eq.~\eqref{eq:plateau}} & Eq.~\eqref{eq:small_pot}  \\ \hline
\multicolumn{1}{|c|}{$\mu$} & \multicolumn{1}{c|}{Eq.~\eqref{eq:mu=1}} & Eq.~\eqref{eq:mu0}  & \multicolumn{1}{c|}{Eq.~\eqref{eq:mu_largephi}} & Eq.~\eqref{eq:large0smalls} \\ \hline
\multicolumn{1}{|c|}{$\zeta$} & \multicolumn{1}{c|}{Eq.~\eqref{eq:eo_small1}} & Eq.~\eqref{eq:eo_small2} & \multicolumn{1}{c|}{Eq.~\eqref{eq:zeta_large1}} & Eq.~\eqref{eq:zeta_large2}  \\ \hline
\end{tabular}
\end{table}

Our model allowed us to derive analytical approximations for a mobility parameter $\mu$ and zeta potential $\zeta$ in different  modes by  suggesting several routes for their tuning. For convenience, we provide a summary of them in Table~\ref{tab:summary1}.

Our results refer to fixed $\phi_0$, so that $\sigma_0$ that obeys
\begin{equation}\label{eq:sigma0}
	\phi _{i}^{\prime }(0)=-\dfrac{4\pi \sigma_0 \ell_{i}}{e},
\end{equation}
where~\cite{vinogradova.oi:2024a}
\begin{equation}
	\phi_i' (0) = -\frac{\sqrt{\gamma}}{ \lambda}\sqrt{2 \left[\cosh \phi_{0} + (\gamma - 1) \cosh \phi_s -\gamma\right]}
\end{equation}
is established self-consistently, so is $\phi_0$ that will become salt-dependent. It would be of interest to extend our analysis to insulators of fixed $\sigma_0$. Note, however, that this will not change general equation \eqref{eq:muf} for $\mu$ we have derived here.

Several of our theoretical predictions could be tested by experiment. An important quantity is the zeta potential because it is feasible to infer it from the  electroosmotic velocity and/or streaming potential that can be measured using available techniques~\cite{kirby.bj:2004,yan.d:2006,werner.c:2001}. It is likely that a fairly detailed comparison between theory and experiment will soon be possible and this should shed more light on the mobility of ions in the Stern layer, as well as, more globally, on the notions about the structure and properties of the electric double layer. Note, however, that while our results can be immediately applied to interpret the velocimetry data, the relation between $\zeta$ and the streaming potential remains an open question that  requires more theoretical studies. 

Another fruitful direction could be to extend calculations based on our double layer model  to planar and cylindrical channels. Existing theories of electro-osmosis  and its repercussion to electrolyte conductivity and energy conversion in thin channels either postulate immobile surface charge~\cite{vanderHeyden.fhj:2006, peters.pb:2016, vinogradova.oi:2023a} or consider $\mu$ below unity~\cite{vinogradova.oi:2021, liu2022, vinogradova.oi:2022}. It is unlikely that a large mobility parameter predicted here will drastically alter the general features of the curves for the channel zeta potential and conductivity, but it will definitely introduce important quantitative changes. The same remark concerns the analysis of electrophoresis of large particles with mobile surface charges~\cite{majhi.s:2024}.

\begin{acknowledgments}
	
	This work was supported by the Ministry of Science and Higher Education of the Russian Federation.
\end{acknowledgments}

\section*{DATA AVAILABILITY}

The data that support the findings of this study are available within the
article.

\section*{AUTHOR DECLARATIONS}

The authors have no conflicts to disclose.


\begin{thebibliography}{46}
\bibitem[not()]{note1}
	Nowadays this obsolete definition of the ``slip plane'' is confusing for the
	newcomers since macroscopically it appears that liquid slips over the plane
	located at distance $\delta +\lambda $ from the wall, but not at the
	$\delta$-plane, where the surface electrostatic potential $\Phi _{s}$ is
	defined.
\end{thebibliography}
\end{document}